\documentclass[12pt]{iopart}

\usepackage{graphics}

\newcommand{\bbox}[1]{\mbox{\boldmath $#1$}}
\newcommand{\beq}{\begin{equation}}
\newcommand{\eeq}{\end{equation}}
\newcommand{\beqa}{\begin{eqnarray}}
\newcommand{\eeqa}{\end{eqnarray}}

\begin{document}

\title{Nuclear forces from chiral EFT: The unfinished business}

\author{R Machleidt$^1$ and D R Entem$^2$}

\address{$^1$ Department of Physics, University of Idaho, Moscow,
Idaho 83844, USA}
\ead{machleid@uidaho.edu}

\address{$^2$ Nuclear Physics Group, University of Salamanca, E-37008 Salamanca, Spain}
\ead{entem@usal.es}

\begin{abstract}
In spite of the great progress we have seen in recent years in the derivation
of nuclear forces from chiral effective field theory (EFT), 
some important issues are still unresolved.
In  this contribution, we discuss the open problems which have particular
relevance for microscopic nuclear structure, namely, the proper renormalization
of chiral nuclear potentials and sub-leading many-body forces.  
\end{abstract}

\maketitle

\section{Introduction}

The fundamental goal of nuclear structure physics is to understand the properties
of atomic nuclei and their reactions in terms of the basic forces between the
constituents. During the past half century, a large variety of phenomenological
forces has been developed and applied in microscopic nuclear structure calculations with some
success. But in the long run phenomenology is not good enough and,
ultimately, we need nuclear interactions that are based upon proper
theory. 
Since the nuclear force is a manifestation of strong interactions, any serious derivation 
has to start from quantum chromodynamics (QCD). 
However, the well-known problem with QCD is that it is non-perturbative 
in the low-energy regime characteristic for nuclear physics.
For many years this fact was perceived as the great obstacle for a derivation
of nuclear forces from QCD---impossible to overcome except by lattice QCD.
The effective field theory (EFT) concept has shown the way out of this dilemma.
One has to realize that the scenario of low-energy QCD is characterized by pions
and nucleons interacting via a force governed by spontaneously broken approximate
chiral symmetry.
This chiral EFT allows for a systematic low-momentum
expansion known as chiral perturbation theory (ChPT)~\cite{Wei79}. 
Contributions are analyzed in terms of powers of small external momenta
over the large scale: $(Q/\Lambda_\chi)^\nu$, where $Q$ is generic for an external
momentum (nucleon three-momentum or pion four-momentum) or pion mass and
$\Lambda_\chi \approx 1$ GeV is the chiral symmetry breaking scale (`hard scale').
The early applications of ChPT focused on systems like $\pi\pi$~\cite{GL84}
and $\pi N$~\cite{GSS88}, where the Goldstone-boson character of the pion
guarantees that the expansion converges.

The past 15 years have also seen great progress in applying ChPT to nuclear forces
\cite{Wei90,Wei92,ORK94,Kol94,Kol99,KBW97,KGW98,EGM98,BK02,EM02,EM02a,EM03,ME05,EGM05,Mac07}.
As a result, nucleon-nucleon ($NN$) potentials of high precision have been constructed, which
are based upon ChPT carried to next-to-next-to-next-to-leading order 
(N$^3$LO)~\cite{EM03,EGM05,Mac07}, and applied in nuclear structure calculations
with great success.

However, in spite of this progress, we are not done. Due to the complexity of the
nuclear force issue, there are still many subtle and not so subtle open problems.
We will not list and discuss all of them, but instead just focus on what is relevant
to nuclear structure physics.  In this regard, there are two important issues
that need our attention:
\begin{itemize}
\item
The proper renormalization of chiral nuclear potentials and
\item
Subleading chiral few-nucleon forces.
\end{itemize}
To set the stage, we will give first, in section~2, a brief overview of chiral nuclear forces and ChPT.
The main two open issues are discussed in section~3 and 4.
The article is concluded in section~5.

\section{Nuclear forces, chiral perturbation theory, and power counting
\label{sec_chpt}}

Effective field theories (EFTs) are defined in terms of effective Langrangians which
are given by an infinite series of terms with increasing number of derivatives
and/or nucleon fields, with the dependence of each term on the pion field 
prescribed by the rules of broken chiral symmetry.
Applying this Lagrangian to a particular process, an unlimited number of Feynman 
graphs can be generated. Therefore,
we need a scheme that makes the theory manageable and calculabel.
This scheme
which tells us how to distinguish between large
(important) and small (unimportant) contributions
is chiral perturbation theory (ChPT), and
determining the power $\nu$ of the expansion
has become known as power counting.

Nuclear potentials are defined as sets of irreducible
graphs up to a given order.
The power $\nu$ of a few-nucleon diagram involving $A$ nucleons
is given in terms of naive dimensional analysis by:
\begin{equation} 
\nu = -2 +2A - 2C + 2L 
+ \sum_i \Delta_i \, ,  
\label{eq_nu} 
\end{equation}
with
\begin{equation}
\Delta_i  \equiv   d_i + \frac{n_i}{2} - 2  \, ,
\label{eq_Deltai}
\end{equation}
where $C$ denotes the number of separately connected pieces and
$L$ the number of loops in the diagram;
$d_i$ is the number of derivatives or pion-mass insertions and $n_i$ the number of nucleon fields 
(nucleon legs) involved in vertex $i$; the sum runs over all vertices contained
in the diagram under consideration.
Note that $\Delta_i \geq 0$
for all interactions allowed by chiral symmetry.
For an irreducible $NN$ diagram (``two-nucleon force'', $A=2$, $C=1$),
(\ref{eq_nu}) collapses to
\begin{equation} 
\nu =  2L + \sum_i \Delta_i \, .  
\label{eq_nunn} 
\end{equation}

The power formula 
(\ref{eq_nu}) 
allows to predict
the leading orders of multi-nucleon forces.
Consider a $m$-nucleon irreducibly connected diagram
($m$-nucleon force) in an A-nucleon system ($m\leq A$).
The number of separately connected pieces is
$C=A-m+1$. Inserting this into
(\ref{eq_nu}) together with $L=0$ and 
$\sum_i \Delta_i=0$ yields
$\nu=2m-4$. Thus, two-nucleon forces ($m=2$) start 
at $\nu=0$, three-nucleon forces ($m=3$) at
$\nu=2$ (but they happen to cancel at that order),
and four-nucleon forces at $\nu=4$ (they don't cancel).
Thus, ChPT provides a straightforward explanation for
the empirically known fact that 2NF $\gg$ 3NF $\gg$ 4NF
\ldots.

In summary, the chief point of the ChPT expansion is that,
at a given order $\nu$, there exists only a finite number
of graphs. This is what makes the theory calculable.
The expression $(Q/\Lambda_\chi)^{\nu+1}$ provides a rough estimate
of the relative size of the contributions left out and, thus,
of the accuracy at order $\nu$.
In this sense, the theory can be calculated to any
desired accuracy and has
predictive power.

Chiral perturbation theory and power counting
imply that nuclear forces emerge as a hierarchy
controlled by the power $\nu$ (see~\cite{ME05} for a pedagogial
introduction).

Since 2003, a very quantitative chiral $NN$ potential (at N$^3$LO,
$\nu=4$)~\cite{EM03}
exists which has been applied successfully in many nuclear structure 
calculations~\cite{Cor02,FOS04,Nav07,Hag08,Bog05}.
However, questions have been raised concerning the proper renormalization
of chiral $NN$ potentials, which is why we will look into this issue in the next section.
Moreover, there are also some open problems in the few-nucleon-force
sector, which will be discussed in section~\ref{sec_manyNF}.

\section{Renormalization of chiral nuclear forces}

\subsection{The chiral $NN$ potential}
In terms of naive dimensional analysis or ``Weinberg counting'',
the various orders of the irreducible graphs which define the chiral $NN$ potential 
are given by:
\beqa
V_{\rm LO} & = & 
V_{\rm ct}^{(0)} + 
V_{1\pi}^{(0)} 
\label{eq_VLO}
\\
V_{\rm NLO} & = & V_{\rm LO} +
V_{\rm ct}^{(2)} + 
V_{1\pi}^{(2)} +
V_{2\pi}^{(2)} 
\label{eq_VNLO}
\\
V_{\rm NNLO} & = & V_{\rm NLO} +
V_{1\pi}^{(3)} + 
V_{2\pi}^{(3)} 
\label{eq_VNNLO}
\\
V_{{\rm N}^3{\rm LO}} & = & V_{\rm NNLO} +
V_{\rm ct}^{(4)} +
V_{1\pi}^{(4)} +  
V_{2\pi}^{(4)} +
V_{3\pi}^{(4)} 
\label{eq_VN3LO}
\eeqa
where 
the superscript denotes the order $\nu$ of the low-momentum
expansion.
LO stands for leading order, NLO for next-to-leading
order, etc..
Contact potentials carry the subscript ``ct'' and
pion-exchange potentials can be identified by an
obvious subscript.

The one-pion exchange (1PE) potential reads
\begin{equation}
V_{1\pi} ({\vec p}~', \vec p) = - 
\frac{g_A^2}{4f_\pi^2}
\: 
\bbox{\tau}_1 \cdot \bbox{\tau}_2 
\:
\frac{
\vec \sigma_1 \cdot \vec q \,\, \vec \sigma_2 \cdot \vec q}
{q^2 + m_\pi^2} 
\,,
\label{eq_1peci}
\end{equation}
where ${\vec p}~'$ and $\vec p$ designate the 
final and initial nucleon momenta 
in the center-of-mass system and
$\vec q \equiv {\vec p}~' - \vec p$  is the 
momentum transfer;
$\vec \sigma_{1,2}$ and $\bbox{\tau}_{1,2}$ are 
the spin and isospin 
operators of nucleon 1 and 2;
$g_A$, $f_\pi$, and $m_\pi$
denote axial-vector coupling constant, the pion decay constant,
and the pion mass, respectively.
Since higher order corrections contribute only to mass and
coupling constant renormalizations and since, on shell,
there are no relativistic corrections, the on-shell
1PE has the form (\ref{eq_1peci}) up to all orders.

Multi-pion exchange, which starts at NLO and continues through
all higher orders, involves
divergent loop integrals that need to be regularized.
An elegant way to do this is dimensional regularization
which 
(besides the main nonpolynomial result) 
typically generates polynomial terms with coefficients
that are, in part, infinite or scale dependent~\cite{KBW97}.
One purpose of the contacts is
to absorb all infinities and scale dependencies and make
sure that the final result is finite and scale independent.
This is the renormalization of the perturbatively calculated
$NN$ amplitude (which, by definition, is the ``$NN$ potential'').
It is very similar to what is done in the ChPT calculations
of $\pi\pi$ and $\pi N$ scattering, namely, a renormalization
order by order, which is the method of choice for any EFT.
Thus, up to this point, the calculation fully meets the
standards of an EFT and there are no problems.
The perturbative $NN$ amplitude can be used to make model
independent predictions for peripheral partial waves~\cite{KBW97,KGW98,EM02a}.

\subsection{Nonperturbative applications of the $NN$ potential}
For calculations of the structure of nuclear few and many-body systems,
the lower partial waves are the most important ones. The fact that
in $S$ waves we have large scattering lengths and shallow (quasi)
bound states indicates that these waves need to be treated nonperturbatively.
Following Weinberg's prescription~\cite{Wei90}, this is accomplished by
inserting the potential $V$ into the Lippmann-Schwinger (LS) equation:
\begin{equation}
 {T}({\vec p}~',{\vec p})= {V}({\vec p}~',{\vec p})+
\int d^3p''\:
{V}({\vec p}~',{\vec p}~'')\:
\frac{M_N}
{{ p}^{2}-{p''}^{2}+i\epsilon}\:
{T}({\vec p}~'',{\vec p}) \,,
\label{eq_LS}
\end{equation}
where $M_N$ denotes the nucleon mass.

In general, the integral in
the LS equation is divergent and needs to be regularized.
One way to do this is  by
multiplying $V$
with a regulator function
\begin{equation}
{ V}(\vec{ p}~',{\vec p}) 
\longmapsto
{ V}(\vec{ p}~',{\vec p})
\;\mbox{\boldmath $e$}^{-(p'/\Lambda)^{2n}}
\;\mbox{\boldmath $e$}^{-(p/\Lambda)^{2n}}
\label{eq_regulator} \,.
\end{equation}
Typical choices for the cutoff parameter $\Lambda$ that
appears in the regulator are 
$\Lambda \approx 0.5 \mbox{ GeV} \ll \Lambda_\chi \approx 1$ GeV.

It is pretty obvious that results for the $T$-matrix may
depend sensitively on the regulator and its cutoff parameter.
This is acceptable if one wishes to build models.
For example, the meson models of the past~\cite{Mac89,MHE87}
always depended sensitively on the choices for the
cutoff parameters which, in fact,
were important for the fit of the $NN$ data.
However, the EFT approach wishes to be fundamental
in nature and not just another model.

In field theories, divergent integrals are not uncommon and methods have
been developed for how to deal with them.
One regulates the integrals and then removes the dependence
on the regularization parameters (scales, cutoffs)
by renormalization. In the end, the theory and its
predictions do not depend on cutoffs
or renormalization scales.

So-called renormalizable quantum field theories, like QED,
have essentially one set of prescriptions 
that takes care of renormalization through all orders. 
In contrast, 
EFTs are renormalized order by order. 

As discussed, the renormalization of {\it perturbative}
EFT calculations is not a problem. {\it The problem
is nonperturbative renormalization.}
This problem typically occurs in {\it nuclear} EFT because
nuclear physics is characterized by bound states which
are nonperturbative in nature.
EFT power counting may be different for nonperturbative processes as
compared to perturbative ones. Such difference may be caused by the infrared
enhancement of the reducible diagrams generated in the LS equation.

Weinberg's implicit assumption~\cite{Wei90,Wei09} was that the counterterms
introduced to renormalize the perturbatively calculated
potential, based upon naive dimensional analysis (``Weinberg counting''),
are also sufficient to renormalize the nonperturbative
resummation of the potential in the LS equation.
In 1996, Kaplan, Savage, and Wise (KSW)~\cite{KSW96}
pointed out that there are problems with the Weinberg scheme
if the LS equation is renormalized 
by minimally-subtracted dimensional regularization.
This criticism resulted in a flurry of publications on
the renormalization of the nonperturbative
$NN$ problem
\cite{FMS00,PBC98,FTT99,Bir06,Bea02,VA05-1,NTK05,VA05-2,VA07,EM06,VA08,Ent08,YEP07,LK08,BKV08,Val08}.
The literature is too comprehensive
to discuss all contributions.
Let us just mention some of the work that has particular relevance
for our present discussion.

If the potential $V$ consists of contact terms only (a.k.a.\
pion-less theory), then
the nonperturbative summation (\ref{eq_LS})
can be performed analytically and the power counting is explicit.
However, when pion exchange is included, then (\ref{eq_LS})
can be solved only numerically and the power counting
is less transparent.
Perturbative ladder diagrams of arbitrarily high order,
where the rungs of the ladder represent a potential made up from
irreducible pion exchange,
suggest that an infinite number of counterterms is needed to achieve
cutoff independence for all the terms of increasing order generated
by the iterations.
For that reason, 
Kaplan, Savage, and Wise (KSW)~\cite{KSW96} proposed 
to sum the leading-order contact interaction to all orders (analytically)
and to add higher-order contacts and
pion exchange perturbatively up to the given order. Unfortunately,
it turned out that the order by order convergence of 1PE 
is poor in the $^3S_1$-$^3D_1$ state~\cite{FMS00}. 
The failure was triggered by the $1/r^3$ singularity of the 1PE tensor
force when iterated to second order. Therefore, KSW counting is no
longer taken into consideration (see, however, \cite{BKV08}).  
A balanced discussion of possible
solutions can be found in \cite{Bea02}.

Some researchers decided to take
a second look at Weinberg's original proposal.
A systematic investigation of Weinberg counting in leading order
has been conducted by Nogga, Timmermans, and van Kolck~\cite{NTK05}
in momentum space, and by Valderrama and Arriola
at LO and higher orders in
configuration space~\cite{VA05-1,VA05-2,VA07}. A comprehensive
discussion of both approaches and their equivalence can be found
in~\cite{Ent08,Val08}.

The LO $NN$ potential is given in (\ref{eq_VLO}) and consists
of 1PE plus two nonderivative contact terms that contribute
only in $S$ waves.
Nogga {\it et al} find that the given counterterms renormalize
the $S$ waves (i.e., stable results are obtained for $\Lambda \rightarrow \infty$) and
the naively expected infinite number of counterterms
is not needed. This means that Weinberg power counting does actually work in
$S$ waves at LO (ignoring the $m_\pi$ dependence of the contact interaction
discussed in Refs.~\cite{KSW96,Bea02}).
However, there are problems with a particular class of higher partial waves,
namely those  
in which the tensor force from 1PE is attractive. The first few cases
of this kind  of low angular momentum are
$^3P_0$, $^3P_2$, and $^3D_2$, which need a counterterm for cutoff independence. 
The leading order (nonderivative) counterterms do not contribute in
$P$ and higher waves, which is why Weinberg counting fails in these cases. 
But the second order contact potential provides counterterms
for $P$ waves. Therefore, the promotion
of, particularly, the $^3P_0$ and $^3P_2$ contacts from NLO to LO would
fix the problem in $P$ waves. To take care of the $^3D_2$ problem,
a N$^3$LO contact, i.e. a term from $V^{(4)}_{\rm ct}$, needs to be promoted to LO.
Partial waves with orbital angular momentum $L\geq 3$ may be calculated in Born
approximation with sufficient accuracy and, therefore, do not pose renormalization
problems.
In this way, one arrives at a scheme
of `modified Weinberg counting'~\cite{NTK05} for the leading order 
two-nucleon interaction.

\subsection{Renormalization beyond leading order}

As discussed, for a quantitative chiral $NN$ potential one needs to advance all the way
to N$^3$LO. Thus, the renormalization issue needs to be discussed beyond LO.
Naively, the most perfect renormalization procedure is the one where the cutoff
parameter $\Lambda$ is carried to infinity while stable results are maintained.
This was done successfully at LO in the work by Nogga {\it et al}~\cite{NTK05} described above.
At NNLO, the infinite-cutoff renormalization procedure has been investigated 
in~\cite{YEP07} for partial waves with total angular momentum $J\leq 1$ and
in~\cite{VA07} for all partial waves with $J\leq 5$. At N$^3$LO, an investigation
of the $^1S_0$ state exists~\cite{Ent08}.
From all of these works, it is evident that no counter term is effective in partial-waves with
short-range repulsion and only a single counter term can effectively be used in
partial-waves with short-range attraction. Thus, for the $\Lambda \rightarrow \infty$
renormalization prescription, even at N$^3$LO, there is either one or no counter term
per partial-wave state. This is inconsistent with any reasonable power-counting scheme
and, therefore, defies the principals of an EFT.

A possible way out of this dilemma was proposed already in~\cite{NTK05}
and reiterated in a recent paper by Long and van Kolck~\cite{LK08}. In the latter
reference, the authors examine the renormalization of an attractive $1/r^2$ potential
perturbed by a $1/r^4$ correction.
Generalizing their findings, they come to
the conclusion that, for any attractive $1/r^n$ potential (with $n\geq 2$),
partial waves with low angular momentum $L$ must be summed to all orders
and one contact term is needed for each $L$ to renormalize the LO
contribution. However, there exists
an angular momentum $L_p$ ($L_p\approx 3$ for the nuclear case, cf.\ ~\cite{NTK05}), 
above which the leading order can be calculated perturbatively. 
In short, naive dimensional analysis (NDA) does not apply at LO below $L_p$.
However, once this failure of NDA is corrected at LO, higher order corrections
can be added in perturbation theory using counterterm that follow NDA~\cite{LK08}.

Reference~\cite{LK08} used just a toy model and, therefore, a full investigation 
using the chiral expansion is needed to answer the question 
if this renormalization approach will work for the realistic nuclear force.
A first calculation of this kind for the $S$ waves was recently performed by
Valderrama~\cite{Val09}. The author renormalizes the LO interaction nonperturbatively
with $\Lambda \rightarrow \infty$ and then uses the LO
distorted wave to calculate the 2PE contributions at NLO and NNLO
perturbatively. It turns out that perturbative renormalizability requires the introduction
of three counterterms in $^1S_0$ and six in the coupled $^3S_1-^3D_1$
channels. 
Thus, the number of counterterms required in this scheme is larger than in the
Weinberg scheme, which reduces the predictive power. For a final evaluation
of this approach, also the results for $P$ and $D$ waves are needed, which
are not yet available.

However, even if such a project turns out to be successful for $NN$ scattering,
there is doubt if the interaction generated in this approach is of any use
for applications in nuclear few- and many-body problems.
In applications, one would first have to solve the many-body problem
with the renormalized LO interaction, and then add higher order corrections in
perturbation theory.
However, it was shown in a recent paper~\cite{Mac09} that the renormalized LO
interaction
is characterized by a very large tensor force from 1PE. This is no surprise since
LO is renormalized with $\Lambda \rightarrow \infty$ implying that the 1PE,
particulary its tensor force, is totally uncut.
As a consequence of this, the wound integral in nuclear matter, $\kappa$,
comes out to be about 40\%. The hole-line and coupled cluster expansions
are know to converge $\sim \kappa^{n-1}$
with $n$ the number of hole-lines or particles per cluster~\cite{Bra67,Day67,Bet71}.
For conventional nuclear forces, the wound integral is typically between 5 and 10\%
and the inclusion of three-body clusters (or three hole-lines) are needed to
obtain converged results in the many-body system~\cite{FOS04,Hag08,DW85}.
Thus, if the wound integral is 40\%, probably, up to six hole-lines need to be
included for approximate convergence. Such calculations are not feasible even with
the most powerful computers of today and will not be feasible any time soon.
Therefore, even if the renormalization procedure proposed in~\cite{LK08} will work
for $NN$ scattering, the interaction produced will be highly impractical (to say
the least) in applications in few- and many-body problems because of convergence problems
with the many-body energy and wave functions.

\subsection{Back to the beginnings}
The various problems with the renormalization procedures discussed above
may have a simple common reason:
An EFT that has validity only for momenta $Q < \Lambda_\chi$ is applied such that
momenta $Q \gg \Lambda_\chi$ are heavily involved (because the regulator cutoff
$\Lambda \rightarrow \infty$).
A recent paper by Epelbaum and Gegelia~\cite{EG09} illustrates the point:
The authors construct an exactly solvable toy-model that simulates a pionful EFT 
and yields finite results for
$\Lambda \rightarrow \infty$.  However, as it turns out, these
finite results are incompatible with the underlying EFT, while
for cutoffs in the order of the hard scale consistency is maintained.
In simple terms, the point to realize is this: 
{\it If an EFT calculation produces (accidentally) a finite result for
$\Lambda \rightarrow \infty$, then that
does not automatically imply that this result is also right.}

This matter is further elucidated in
the lectures by Lepage of 1997~\cite{Lep97}.
Lepage points out that it makes little sense to take the momentum cutoff beyond
the range of validity of the effective theory. By assumption, our data involves energies
that are too low---wave lengths that are too long---to probe the true structure of
the theory at very short distances. When one goes beyond the hard-scale of the theory,
structures are seen that are almost certainly wrong. Thus, results cannot improve
and, in fact, they may degrade or, in more extreme cases, the theory may become
unstable or untunable. In fact, in the $NN$ case, this is what is happening in 
several partial waves (as reported above). Therefore, Lepage
suggests to take the following three steps when building an effective theory:
\begin{enumerate}
\item
Incorporate the correct long-range behavior: The long-range behavior of the underlying
theory must be known, and it must be built into the effective theory. In the case of
nuclear forces, the long-range theory is, of course, well known and 
given by one- and multi-pion exchanges.
\item
Introduce an ultraviolet cutoff to exclude high-momentum states, or, equivalent, to soften the 
short-distance behavior: The cutoff has two effects: First it excludes high-momentum states,
which are sensitive to the unknown short-distance dynamics; only states that we understand
are retained. Second it makes all interactions regular at $r=0$, thereby avoiding the infinities
that beset the naive approach.
\item
Add local correction terms (also known as contact or counter terms) 
to the effective Hamiltonian. These mimic the effects of the 
high-momentum states excluded by the cutoff introduced in the previous step.
In the meson-exchange picture, the short-range nuclear force is described by
heavy meson exchange, like the $\rho(770)$ and $\omega(782)$. However, at low
energy, such structures are not resolved. Since we must include contact terms 
anyhow, it is most efficient to use them to account for any heavy-meson exchange
as well.
The correction terms systematically remove dependence on the cutoff.
\end{enumerate}

A first investigation in the above spirit has been
conducted by Epelbaum and Mei\ss ner~\cite{EM06} in 2006.
The authors stress that there is no point in taking the cutoff $\Lambda$ beyond the
breakdown scale of the EFT, $\Lambda_\chi \approx m_\rho \approx 1$ GeV,
since the error of the calculation is not expected to decrease
in that regime.
Any value for the cutoff parameter $\Lambda$
is acceptable if the error associated with its finite value is within
the theoretical uncertainty at the given order.
The authors conduct an investigation at LO (including only the counter terms
implied by Weinberg counting)
and find that,
starting from $\Lambda \approx 3$ fm$^{-1}$, the error
in the $NN$ phase shifts due to keeping $\Lambda$ finite
stays within the theoretical uncertainty at LO.

\subsection{Bringing the renormalization business to a finish}

Crucial for an EFT are regulator independence (within the range of validity
of the EFT) and a power counting scheme that allows for order-by-order
improvement with decreasing truncation error.
The purpose of renormalization is to achieve this regulator independence while maintaining
a functional power counting scheme.
After the comprehensive tries and errors of the past, it appears that there are two renormalization
schemes which have the potential to achieve the above goals and, therefore, should be investigated 
systematically in the near future---with the hope to bring the tiresome renormalization
issue finally to conclusion.

In {\it scheme one}, the LO calculation is conducted nonperturbatively
(with $\Lambda \rightarrow \infty$ as in~\cite{NTK05})
and subleading orders are added perturbatively in distorted wave Born approximation.
As mentioned above,
Valderrama has started this in $S$ waves~\cite{Val09}, but results in higher
partial waves are needed to fully assess this approach.
Even though at this early stage any judgement is speculative, we take the liberty to predict
that this approach will be only of limited success and utility---for the following reasons.
First, it will probably require about twice as many counterterms as Weinberg counting
and, therefore, will have less predictive power. Second, this scheme may converge badly,
because the largest portion of the nuclear force, namely, the intermediate-range
attraction appears at NNLO. Third, as discussed in~\cite{Mac09}, this force may be problematic
(and, therefore, impractical) in applications in nuclear few- and many-body systems, 
because of a pathologically strong tensor force that will cause bad convergence of energy and wave functions. Finally, in the work that has been conducted so far within this
scheme by Valderrama,
it is found that only rather soft cutoffs can be used.

The latter point (namely, soft cutoffs) suggests that one may then as well
conduct the calculation nonperturbatively at all orders (up to N$^3$LO) using Weinberg counting, 
which is no problem with soft cutoffs. This is {\it scheme two} that we propose to
investigate systematically. 
In the spirit of Lepage, the cutoff independence should be examined
for cutoffs below the hard scale and not beyond. Ranges of cutoff independence within the
theoretical error are to be identified using `Lepage plots'~\cite{Lep97}.
A very systematic investigation of this kind does not exist at this time and is therefore needed,
once and for all and for principal reasons.
Comprehensive circumstantial
evidence from the numerous chiral $NN$ potentials constructed over the past 
decade~\cite{ORK94,EGM98,EM02,EM03,ME05,EGM05}
may be perceived as an indication that this investigation will most likely be a success. 
If so, then the results of such investigation will 
resolve the renormalization issue
and remove concerns about
existing chiral $NN$ potentials~\cite{EM03,EGM05},
which are currently applied in
microscopic nuclear structure physics with great success.

\section{Few-nucleon forces and what is missing \label{sec_manyNF}}

We will now discuss the other issue we perceive as unfinished and important, namely,
subleading chiral few-nucleon forces.

Nuclear three-body forces in ChPT were initially discussed
by Weinberg~\cite{Wei92}.
The 3NF at NNLO, was derived by van Kolck~\cite{Kol94}
and applied, for the first time, in nucleon-deuteron
scattering by Epelbaum {\it et al}~\cite{Epe02b}.
The leading 4NF (at N$^3$LO) was recently constructed by
Epelbaum~\cite{Epe06} and found to contribute in the
order of 0.1 MeV to the $^4$He binding energy
(total $^4$He binding energy: 28.3 MeV)
in a preliminary calculation~\cite{Roz06}, confirming the traditional
assumption that 4NF are essentially negligible.
{\bf Therefore, the focus is on 3NF.}

For a 3NF, we have $A=3$ and $C=1$ and, thus, (\ref{eq_nu})
implies for 3NF
\begin{equation}
\nu = 2 + 2L + 
\sum_i \Delta_i \,.
\label{eq_nu3nf}
\end{equation}
We will use this equation to analyze 3NF contributions
order by order.
The lowest possible power is obviously $\nu=2$ (NLO), which
is obtained for no loops ($L=0$) and 
only leading vertices
($\sum_i \Delta_i = 0$). 
This 3NF happens to vanish~\cite{Wei92,YG86,CF86}.
The first non-vanishing 3NF occurs at NNLO.

\begin{figure}[t]\centering
\vspace*{-4.5cm}
\hspace*{-2.0cm}
\scalebox{0.65}{\includegraphics{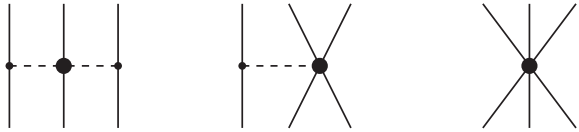}}
\vspace*{-11.7cm}
\caption{The three-nucleon force at NNLO.
From left to right: 2PE, 1PE, and contact diagrams.
Solid lines represent nucleons and dashed lines pions.
Small dots denote vertices with $\Delta_i = 0$ and large solid dots are $\Delta_i=1$.}
\label{fig_3nf_nnlo}
\end{figure}

\subsection{The 3NF at NNLO}
The power $\nu=3$ (NNLO) is obtained when
there are no loops ($L=0$) and 
$\sum_i \Delta_i = 1$, i.e., 
$\Delta_i=1$ for one vertex 
while $\Delta_i=0$ for all other vertices.
There are three topologies which fulfill this condition,
known as the two-pion exchange (2PE), 1PE,
and contact graphs
(figure~\ref{fig_3nf_nnlo}).

The 3NF at NNLO
has been derived (without the $1/M_N$ corrections)~\cite{Kol94,Epe02b}
and applied in
calculations of few-nucleon reactions~\cite{Epe02b,KE07,Viv08,Mar09},
structure of light- and medium-mass 
nuclei~\cite{Nav07,Hag07,Ots09},
and nuclear and neutron matter~\cite{Bog05,HS09}
with a fair deal of success.
However, the famous `$A_y$ puzzle' of nucleon-deuteron scattering
is not solved~\cite{Epe02b,KE07}, and the even bigger problem with the
analyzing power in $p$-$^3$He scattering~\cite{Fis06,DF07} will certainly not be fixed
at this order.
Furthermore, the spectra of light nuclei leave room for improvement~\cite{Nav07}.

We note that there are further 3NF contributions at NNLO, namely, the
$1/M_N$ corrections of the NLO 3NF diagrams.
Part of these corrections 
have been calculated by Coon and Friar in 1986~\cite{CF86}.
These contributions are believed to be very small.

In summary, because of various unresolved problems in microscopic nuclear structure, 
the 3NF beyond NNLO is very much in need. In fact, it is no exaggeration to state that
the 3NF at sub-leading orders
is presently one of the most important outstanding issues in the chiral EFT
approach to nuclear forces.

\begin{figure}[t]\centering
\vspace*{-4.8cm}
\hspace*{-1.5cm}
\scalebox{0.70}{\includegraphics{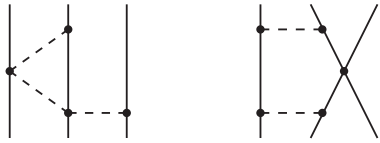}}
\vspace*{-12.5cm}
\caption{The 3NF at N$^3$LO:
Two examples of one-loop graphs.
Notation as in figure~\ref{fig_3nf_nnlo}.}
\label{fig_3nf_n3lo}
\end{figure}

\subsection{The 3NF at N$^3$LO}

According to (\ref{eq_nu3nf}),
the value $\nu=4$, which corresponds to N$^3$LO, is obtained
for the following classes of diagrams.

\subsubsection*{3NF loop diagrams at N$^3$LO.}
For this group of graphs, we have
$L=1$ and, therefore, all $\Delta_i$ have to be zero
to ensure $\nu=4$. 
Thus, these one-loop 3NF diagrams can include
only leading order vertices, the parameters of which
are fixed from $\pi N$ and $NN$ analysis.
We show two samples
of this very large class of diagrams
in figure~\ref{fig_3nf_n3lo}. 
One sub-group of these diagrams (``$2\pi$ exchange graphs'')
has been calculated by Ishikawa and Robilotta~\cite{IR07},
and two other topologies ($2\pi$-$1\pi$ and ring diagrams)
have been evaluated by the Bonn-J\"ulich group~\cite{Ber08}.
The remaining topologies, which involve a leading order four-nucleon
contact term (e.g., second diagram of figure~\ref{fig_3nf_n3lo}),
are under construction by the Bonn-J\"ulich group.
The N$^3$LO $2\pi$-exchange 3NF has been applied in the calculation
of nucleon-deuteron observables in ~\cite{IR07} 
producing very small effects.

The smallness of the $2\pi$ loop 3NF at N$^3$LO is not unexpected.
It is consistent with
experience with corresponding 2NF diagrams: 
the NLO 2PE contribution to the $NN$ potential, which 
involves one loop and only leading vertices, is also relatively small.

By the same token, one may expect that also all the other N$^3$LO
3NF loop topologies will produce only small effects.

\subsubsection*{3NF tree diagrams at N$^3$LO.}
The order $\nu=4$ is also obtained for the combination $L=0$ (no loops)
and $\sum_i \Delta_i = 2$.
Thus, either two vertices have to carry $\Delta_i=1$ or
one vertex has to be of the $\Delta_i=2$ kind,
while all other vertices are $\Delta_i=0$.
This is achieved if 
in the NNLO 3NF graphs of figure~\ref{fig_3nf_nnlo}
the power of one vertex is raised by one.
The latter happens if a relativistic
$1/M_N$ correction is applied.
A closer inspection reveals that all $1/M_N$ corrections of the
NNLO 3NF vanish and the first non-vanishing corrections
are proportional to $1/M_N^2$ and appear at N$^4$LO.
However, there are non-vanishing $1/M_N^2$ corrections of the NLO 3NF
and there are so-called drift corrections~\cite{Rob06} 
which contribute at N$^3$LO (some drift corrections are claimed to
contribute even at NLO~\cite{Rob06}). 
We do not expect these contributions to be sizable.
Moreover, there are contributions from the $\Delta_i =2$
Lagrangian~\cite{FMS98} proportional to the
low-energy constants $d_i$. As it turns out, these terms have
at least one time-derivative, which causes them to be
$Q/M_N$ suppressed and demoted to N$^4$LO.

Thus, besides some minor $1/M_N^2$ corrections, there are no tree
contributions to the 3NF at N$^3$LO.

\subsubsection*{Summarizing the entire N$^3$LO 3NF contribution:}
For the reasons discussed, we anticipate that this 3NF is weak and will not solve
any of the outstanding problems. 
In view of this expectation, we have to look for
more sizable 3NF contributions elsewhere.

\subsection{The 3NF at N$^4$LO of the $\Delta$-less theory}
The obvious step to take is to proceed to the next order,
N$^4$LO or $\nu=5$, of the $\Delta$-less theory which is the one
we have silently assumed so far. (The $\Delta$-full theory will
be introduced and discussed below.)
Some of the tree diagrams that appear at this order were mentioned already:
the $1/M_N^2$ corrections of the NNLO 3NF and the trees with one $d_i$
vertex which are $1/M_N$ suppressed. Because of the suppression factors,
we do not expect sizable effects from these graphs.
Moreover, there are also tree diagrams with one vertex from the
$\Delta_i=3$ $\pi N$ Lagrangian~\cite{Fet00,FM00} proportional to the 
LECs $e_i$. Because of the high dimension of these vertices and assuming
reasonable convergence, we do not anticipate much from these trees either.

\begin{figure}[t]\centering
\vspace*{-1.5cm}
\hspace*{-5.0cm}
\scalebox{0.70}{\includegraphics{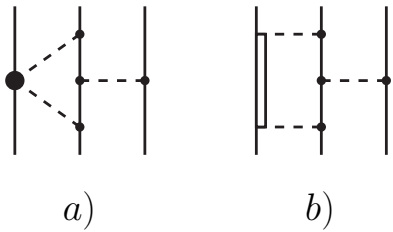}}
\vspace*{-15.00cm}
\caption{(a) One-loop 3NF at N$^4$LO of the $\Delta$-less theory.
(b) Corresponding diagram of the $\Delta$-full theory which contributes
at N$^3$LO.
Double lines represent $\Delta$ isobars; other
notation as in figure~\ref{fig_3nf_nnlo}.}
\label{fig_3nf_n4lo}
\end{figure}

However, we believe that the loop contributions that occur at this order are truly important.
They are obtained by replacing in the N$^3$LO loops (figure~\ref{fig_3nf_n3lo})
one vertex by a $\Delta_i=1$ vertex [with LEC $c_i$]. 
We show one symbolic example of this large group of diagrams
in figure~\ref{fig_3nf_n4lo}(a).
This 3NF is presumably large and, thus, what we are looking for.

The reasons, why these graphs are large, can be argued as follows.
Corresponding 2NF diagrams are the three-pion exchange (3PE)
contributions to the $NN$ interaction. In analogy to 
Figs.~\ref{fig_3nf_n3lo} and \ref{fig_3nf_n4lo}(a),
there are 3PE 2NF diagrams with only leading vertices and the ones with one (sub-leading) 
$c_i$ vertex (and the rest leading). These diagrams have been evaluated by Kaiser in 
Refs.~\cite{Kai00} and \cite{Kai01}, respectively. 
Kaiser finds that the 3PE contributions with one sub-leading vertex are about
an order magnitude larger then the leading ones.

\subsection{N$^3$LO 3NF contributions in the $\Delta$-full theory}

The above considerations indicate that the $\Delta$-less theory exhibits,
in some cases, a bad convergence pattern.
The reason for the unnaturally strong subleading contributions are the large
values of the $\Delta_i=1$ LECs, $c_i$. The large values can be explained
in terms of resonance saturation~\cite{BKM97}.
The $\Delta(1232)$-resonance contributes considerably to $c_3$ and $c_4$.
The explicit inclusion of the $\Delta$ takes strength out of these
LECs and moves this strength to a lower order, thus improving the
convergence~\cite{OK92,ORK94,KGW98,KEM07,EKM08}. 
Figure~\ref{fig_3nf_n4lo} illustrates this fact for the 3NF
under consideration: 
the diagram of the $\Delta$-less theory shown in (a)
is (largely) equivalent to diagram (b) which includes one $\Delta$ excitation.
Note, however,
that diagram (a) is N$^4$LO, while diagram (b) is N$^3$LO.
Moreover, there are further N$^3$LO one-loop diagrams with two and three $\Delta$ excitations,
which correspond to diagrams of order N$^5$LO and N$^6$LO, respectively, in the $\Delta$-less
theory. 

This consideration clearly shows that the inclusion of $\Delta$ degrees
of freedom in chiral EFT makes the calculation of sizable higher-order 3NF 
contributions much more efficient.

\subsection{Summarizing the open 3NF business}

To make a complicated story short, this is the bottom line concerning 3NF:
\begin{itemize}
\item
The chiral 3NF at NNLO is insufficient. Additional {\it sizable}
3NF contributions are needed.
\item
The chiral 3NF at N$^3$LO (in the $\Delta$-less theory) 
most likely does {\it not} produce sizable contributions.
\item
Sizable contributions are expected from one-loop 3NF diagrams
at N$^4$LO of the $\Delta$-less or N$^3$LO of the $\Delta$-full
theory (figure~\ref{fig_3nf_n4lo}). {\it These 3NF contributions may
turn out to be the missing pieces in the 3NF puzzle and have the potential to solve
the outstanding problems in microscopic nuclear structure.}\footnote{Note that the Illinois
3NF~\cite{Pie01} includes two one-loop diagrams with one and two $\Delta(1232)$-isobars.
The deeper reason for this may be in arguments we are presenting.}
\end{itemize}

\section{Conclusions and Outlook}

The past 15 years have seen great progress in our understanding of nuclear forces
in terms of low-energy QCD. Key to this development was the realization that
low-energy QCD is equivalent to an effective field theory (EFT) which allows for 
a perturbative expansion that has become know as chiral perturbation theory (ChPT).
In this framework, two- and many-body forces emerge on an equal footing and the empirical fact
that nuclear many-body forces are substantially weaker than the two-nucleon force
is explained automatically.

In spite of the great progress and success of the past 15 years, there are still some
unresolved issues that will need our attention in the near future. One problem is the
proper renormalization of the chiral two- and many-nucleon potentials. 
We have discussed this issue in section~3, where we also spelled out the systematic work
that needs to be done to resolve the problems.

The other unfinished business are the few-nucleon forces beyond NNLO (``sub-leading
few-nucleon forces'') which are needed to hopefully resolve some important outstanding
nuclear structure problems. We believe that we identified correctly where these
forces will emerge within the systematic scheme of ChPT.

If the open issues discussed in this paper will be resolved within
the next few years, then, after 80 years of desperate struggle, we
may finally claim that the nuclear force problem is essentially solved.
The greatest beneficiary of such progress will be the field of {\it ab initio} nuclear
structure physics.

\section*{Acknowledgments}
This work was supported in part by 
the US Department of Energy under
Grant No.\ DE-FG02-03ER41270.


\section*{References}
\begin{thebibliography}{99}
\bibitem{Wei79} Weinberg S 1979 {\it Physica} {\bf 96A} 327
\bibitem{GL84}
Gasser J and Leutwyler H 1984
{\it Ann. Phys. (N.Y.)} {\bf 158} 142
\bibitem{GSS88}
Gasser J Sainio M E and \v{S}varc A 1988
{\it Nucl. Phys.} {\bf B307} 779
\bibitem{Wei90} 
Weinberg S 1990 {\it Phys.\ Lett.} B {\bf 251} 288
\nonum
Weinberg S 1991 {\it Nucl.\ Phys.} {\bf B363} 3
\bibitem{Wei92} Weinberg S 1992
{\it Phys.\ Lett.} B {\bf 295} 114
\bibitem{ORK94}
Ord\'o\~nez C {\it et al} 1994
{\it Phys.\ Rev.\ Lett.} {\bf 72} 1982
\nonum
Ord\'o\~nez C {\it et al} 1996
{\it Phys.\ Rev.} C {\bf 53} 2086
\bibitem{Kol94} van Kolck U 1994
{\it Phys. Rev.} C {\bf 49} 2932
\bibitem{Kol99} van Kolck U 1999
{\it Prog.\ Part.\ Nucl.\ Phys.} {\bf 43} 337
\bibitem{KBW97} Kaiser N, Brockmann R and Weise W 1997
{\it Nucl.\ Phys.} {\bf A625} 758
\bibitem{KGW98} Kaiser N, Gerstend\"orfer S and Weise W 1998
{\it Nucl.\ Phys.} {\bf A637} 395
\bibitem{EGM98} 
Epelbaum E {\it et al} 1998
{\it Nucl.\ Phys.} {\bf A637} 107
\nonum
Epelbaum E {\it et al} 2000
{\it Nucl. Phys.} {\bf A671} 295
\bibitem{BK02} Bedaque P F and van Kolck U 2002 
{\it Ann. Rev. Nucl. Part. Sci.} {\bf 52} 339
\bibitem{EM02} Entem D R and Machleidt R 2002
{\it Phys. Lett.} B {\bf 524} 93
\bibitem{EM02a} Entem D R and Machleidt R 2002
{\it Phys. Rev.} C {\bf 66} 014002
\bibitem{EM03} Entem D R and Machleidt R 2003
{\it Phys. Rev.} C {\bf 68} 041001
\bibitem{ME05} Machleidt R and Entem D R 2005
{\it J. Phys. G: Nucl. Part. Phys.} {\bf 31} S1235
\bibitem{EGM05} Epelbaum E, Gl\"ockle W and Mei\ss ner U-G 2005
{\it Nucl. Phys.} {\bf A747} 362
\bibitem{Mac07} Machleidt R 2007
Nuclear forces from chiral effective field theory
{\it Preprint} arXiv:0704.0807 [nucl-th]
\bibitem{Cor02} 
Coraggio L {\it et al} 2002 {\it Phys. Rev.} C {\bf 66} 021303
\nonum
Coraggio L {\it et al} 2005 {\it Phys. Rev.} C {\bf 71} 014307
\bibitem{FOS04}
Fujii S {\it et al} 2004 {\it Phys. Rev.} C {\bf 69} 034328
\nonum
Fujii S {\it et al} 2009 {\it Phys. Rev. Lett.} {\bf 103} 182501
\bibitem{Nav07} Navratil P {\it et al} 2007 {\it Phys. Rev. Lett.} {\bf 99} 042501
\bibitem{Hag08} 
Hagen G {\it et al} 2008 {\it Phys. Rev. Lett.} {\bf 101} 092502
\nonum
Hagen G {\it et al} 2009 {\it Phys. Rev.} C {\bf 80} 021306
\bibitem{Bog05} 
Bogner S K {\it et al} 2005 {\it Nucl. Phys.} {\bf A763} 59
\nonum
Bogner S K {\it et al} 2009 Nuclear matter from chiral low-momentum interactions
{\it Preprint} arXiv:0903.3366 [nucl-th]
\bibitem{Mac89} Machleidt R 1989
{\it Adv. Nucl. Phys.} {\bf 19} 189
\bibitem{MHE87}
Machleidt R, Holinde K and Elster Ch 1987
{\it Phys.\ Rep.} {\bf 149} 1
\bibitem{Wei09}
Weinberg S 2009 Effective Field Theory, Past and Future {\it Preprint}
arXiv:0908.1964 [hep-th]
\bibitem{KSW96}
Kaplan D B, Savage M J and Wise M B 1996 {\it Nucl. Phys.} {\bf B478} 629
\nonum
Kaplan D B, Savage M J and Wise M B 1998 {\it Phys. Lett.} B {\bf 424} 390
\nonum
Kaplan D B, Savage M J and Wise M B 1998 {\it Nucl. Phys.} {\bf B534} 329
\bibitem{FMS00}
Fleming S, Mehen T and Stewart I W 2000 {\it Nucl. Phys.} {\bf A677} 313
\nonum
Fleming S, Mehen T and Stewart I W 2000 {\it Phys. Rev.} C {\bf 61} 044005
\bibitem{PBC98}
Phillips D R, Beane S R and Cohen T D 1998 {\it Ann. Phys. (N.Y.)} {\bf 263} 255
\bibitem{FTT99}
Frederico T, Timoteo V S and Tomio L 1999 {\it Nucl. Phys.} {\bf A653} 209
\bibitem{Bir06}
M. C. Birse M C 2006 {\it Phys. Rev.} C {\bf 74} 014003
\nonum
M. C. Birse M C 2007 {\it Phys. Rev.} C {\bf 76} 034002
\bibitem{Bea02} 
Beane S R, Bedaque P F, Savage M J and van Kolck U 2002
{\it Nucl. Phys.} {\bf A700} 377
\bibitem{VA05-1}
Pavon Valderrama M and Ruiz Arriola E 2005
{\it Phys.\ Rev.}  C {\bf 72}  054002
\bibitem{NTK05}
Nogga A, Timmermans R G E and van Kolck U 2005
{\it Phys. Rev.} C {\bf 72} 054006
\bibitem{VA05-2}
Pavon Valderrama M and Ruiz~Arriola E 2006
{\it Phys.\ Rev.}  C {\bf 74}  054001 
\bibitem{VA07}
Pavon Valderrama M and Ruiz Arriola E 2006
{\it Phys.\ Rev.}  C {\bf 74}  064004,
Erratum 2007 {\it Phys. Rev.} C {\bf 75} 059905
\bibitem{EM06}
Epelbaum E and Mei\ss ner U-G 2006
On the renormalization of the one-pion exchange potential and the consistency of
Weinberg's power counting
{\it Preprint} arXiv:nucl-th/0609037
\bibitem{VA08}
Pavon Valderrama M and Ruiz Arriola E 2008
{\it Ann. Phys. (N.Y.)} {\bf 323} 1037
\bibitem{Ent08} Entem D R, Ruiz Arriola E, Pav\'on Valderrama M and Machleidt R 2008
{\it Phys. Rev.} C {\bf 77} 044006
\bibitem{YEP07}
Yang C-J, Elster Ch and Phillips D R 2008 {\it Phys. Rev.} C {\bf 77} 014002
\nonum
Yang C-J, Elster Ch and Phillips D R 2009 {\it Phys. Rev.} C {\bf 80} 034002
\nonum
Yang C-J, Elster Ch and Phillips D R 2009 {\it Phys. Rev.} C {\bf 80} 044002
\bibitem{LK08} Long B and van Kolck U 2008 
{\it Ann. Phys. (N.Y)} {\bf 323} 1304
\bibitem{BKV08} Beane S R, Kaplan D B and Vuorinen A 2008
Perturbative nuclear physics
{\it Preprint} arXiv:0812.3938 [nucl-th]
\bibitem{Val08}
Pavon Valderrama M, Nogga A, Ruiz Arriola E and Phillips D R 2008
{\it Eur.\ Phys.\ J.}  A {\bf 36} 315
\bibitem{Val09}
Valderrama M P 2009
Perturbative Renormalizability of Chiral Two Pion Exchange in Nucleon-Nucleon
Scattering {\it Preprint} arXiv:0912.0699 [nucl-th]
\bibitem{Mac09}
Machleidt R, Liu P, Entem D R and Arriola E R 2009
Renormalization of the leading-order chiral nucleon-nucleon interaction
and bulk properties of nuclear matter 
{\it Preprint} arXiv:0910.3942 [nucl-th]
\bibitem{Bra67}
Brandow B H 1967 {\it Rev. Mod. Phys.} {\bf 39} 771
\bibitem{Day67}
Day B D 1967 {\it Rev. Mod. Phys.} {\bf 39} 719
\bibitem{Bet71}
Bethe H A 1971 {\it Ann. Rev. Nucl. Sci.} {\bf 21} 93
\bibitem{DW85}
Day B D and Wiringa R B 1985 {\it Phys. Rev.} C {\bf 32} 1057
\bibitem{EG09}
Epelbaum E and Gegelia J 2009 
Regularization, renormalization and ``peratization'' in effective field theory for two nucleons 
{\it Preprint} arXiv:0906.3822 [nucl-th]
\bibitem{Lep97}
Lepage G P 1997 How to Renormalize the Schr\"odinger Equation
{\it Preprint} nucl-th/9706029
\bibitem{Epe02b}
Epelbaum E {\it et al} 2002
{\it Phys. Rev.} C {\bf 66} 064001
\bibitem{Epe06}
Epelbaum E 2006 {\it Phys.\ Lett.} B {\bf 639} 456
\nonum
Epelbaum E 2007 {\it Eur. Phys. J.} {\bf A34} 197
\bibitem{Roz06}
Rozpedzik D {\it et al} 2006 
{\it Acta Phys. Polon.} {\bf B37} 2889 {\it Preprint} arXiv:nucl-th/0606017
\bibitem{YG86} Yang S N and Gl\"ockle W 1986
{\it Phys. Rev.} C {\bf 33} 1774
\bibitem{CF86}
Coon S A and Friar J L 1986 
{\it Phys. Rev.} C {\bf 34} 1060
\bibitem{KE07}
Kalantar-Nayestanaki N {\it et al} 2007 
The three-nucleon system as a laboratory for nuclear physics:
the need for 3N forces
{\it Preprint} arXiv:nucl-th/0703089,
and references therein.
\bibitem{Viv08}
Viviani M, Kievsky A, Girlanda L and Marcucci L C 2009
{\it Few Body Syst.} {\bf 45} 119
\bibitem{Mar09}
Marcucci L C, Kievsky A, Girlanda L, Rosati S and Viviani M 2009
{\it Phys. Rev.} C {\bf 80} 034003
\bibitem{Hag07}
Hagen G {\it et al} 2007 {\it Phys. Rev.} C {\bf 76} 034302
\bibitem{Ots09}
Otsuka T {\it et al} 2009
Three-body forces and the limit of oxygen isotopes
{\it Preprint} arXiv:0908.2607
\bibitem{HS09}
Hebeler K and Schwenk A 2009
Chiral three-nucleon forces and neutron matter
{\it Preprint} arXiv:0911.0483 [nucl-th]
\bibitem{Fis06}
Fisher B M {\it et al} 2006
{\it Phys. Rev.} C {\bf 74} 034001
\bibitem{DF07}
Deltuva A and Fonseca A C 2007
{\it Phys. Rev. Lett.} {\bf 98} 162502
\bibitem{IR07} 
Ishikawa S and Robilotta M R 2007 
{\it Phys. Rev.} C {\bf 76} 014006
\bibitem{Ber08}
Bernard V, Epelbaum E, Krebs H, and Mei\ss ner U-G 2008
{\it Phys. Rev.} C {\bf 77} 064004
\bibitem{Rob06}
Robilotta M R 2006 
{\it Phys. Rev.} C {\bf 74} 044002
\bibitem{FMS98}
Fettes N, Mei\ss ner U-G and Steininger S 1998
{\it Nucl. Phys.} {\bf A640} 199
\bibitem{Fet00}
Fettes N {\it et al} 2000 {\it Ann. Phys. (N.Y.)} {\bf 283} 273,
Erratum 2001 {\it Ann. Phys. (N.Y.)} {\bf 288} 249
\bibitem{FM00}
Fettes N and Mei\ss ner U-G 2000 
{\it Nucl. Phys.} {\bf A676} 311
\bibitem{Kai00}
Kaiser N 2000 {\it Phys. Rev.} C {\bf 61} 014003
\nonum
Kaiser N 2000 {\it Phys. Rev.} C {\bf 62} 024001
\bibitem{Kai01}
Kaiser N 2001 {\it Phys. Rev.} C {\bf 63} 044010
\bibitem{BKM97}
Bernard V, Kaiser N, and Mei\ss ner U-G 1997
{\it Nucl. Phys.} {\bf A615} 483
\bibitem{OK92}
Ord\'o\~nez C and van Kolck U 1992
{\it Phys.\ Lett.} B {\bf 291} 459
\bibitem{KEM07}
Krebs H, Epelbaum E and Mei\ss ner U-G 2007
{\it Eur. Phys. J.} {\bf A32} 127
\bibitem{EKM08}
Krebs H, Epelbaum E and Mei\ss ner U-G 2008
{\it Nucl. Phys.} {\bf A806} 65
\bibitem{Pie01}
Pieper S C {\it et al} 2001
{\it Phys. Rev.} C {\bf 64} 014001
\end{thebibliography}
\end{document}